\documentclass[amsmath,amssymb,aps,prl,reprint,twocolumn]{revtex4-2}
\usepackage{bm}
\usepackage[rgb]{xcolor}
\usepackage{graphicx}
\usepackage{physics}
\usepackage{comment}
\usepackage{enumitem}
\usepackage{hyperref}
\usepackage[capitalize]{cleveref}

\graphicspath{{figures/}{.}}

\begin{document}

\title{Conservative yet constitutively odd elasticity in prestressed metamaterials}

\author{Tyler A. Engstrom}
\email{tyler.engstrom@unco.edu} 
\affiliation{Department of Physics and Astronomy, University of Northern Colorado, Greeley, CO 80639}

\author{Daniel M. Sussman}
\email{daniel.m.sussman@emory.edu} 
\affiliation{Department of Physics, Emory University, Atlanta, GA 30322}

\begin{abstract}
    We introduce a design principle for mechanical metamaterials based on ``odd elasticity, once removed.''
    By revisiting classic results relating the variation of Cauchy stress and Lagrangian strain around a prestressed reference state, we show how anisotropic, equilibrium prestress can generate a major anti-symmetry in the material's constitutive response.
    Tuning the system to such a state drives it to a critical instability, radically transforming its acoustic properties.
    We demonstrate this by inverse-designing uniform 2D solids that act as unique waveguides supporting decoupled modes along special lattice directions: a string-like mode and an exotic soft mode that is in-plane but has flexural character ($\omega \sim q^2$) and exhibits a remarkable ``DC'' energy current.
    The magnitude of the anisotropic prestress acts as a control parameter for small-$q$ stability, and, when taken to zero, transforms the string-like mode into another soft mode, giving support for circularly polarized ``spin'' waves with arbitrary $q$. 
    This principle of harnessing conservative ``oddness'' to unlock instability-driven wave phenomena provides a powerful new route to creating tunable materials for guiding and controlling mechanical waves. 
\end{abstract}

\maketitle


This work bridges two vibrant areas of study in modern mechanics.
The first, mechanical metamaterials, leverages carefully designed microstructures to sculpt specific mechanical responses (ranging from controlled wave propagation to unusual bulk elasticity) \cite{florijn2014programmable,craster2023mechanical,surjadi2019mechanical,czajkowski2022conformal}.
The second, odd elasticity \cite{scheibner2020odd,surowka2023odd,fruchart2023odd,fossati2024odd,shankar2024active,lee2025odd}, explores materials with non-conservative internal forces.
The defining characteristic of these ``odd'' solids is an elastic tensor with major anti-symmetry, $K_{ijkl} = -K_{klij}$.
This anti-symmetry originates in active internal processes that break energy conservation, leading to exotic and non-reciprocal phenomena that cannot be found in conventional materials.

Can the mathematical hallmark of odd elasticity -- a major anti-symmetry in an elastic response tensor -- be realized in a purely conservative system? If so, what are the physical consequences of this?
We revisit the classic theory of elasticity around a prestressed reference state \cite{barron1965second,wallace1967thermoelasticity} and demonstrate that this major anti-symmetry can emerge, not in the fundamental elastic tensor governing the potential energy, but in the effective tensor relating the variation of Cauchy stress to Lagrangian strain.
This is done via carefully tuned anisotropic prestress, allowing the system's energy landscape itself to generate a response that is constitutively odd but energetically even.
Given its anti-symmetric structure and non-energy-conjugate matchmaking (it relates a stress and strain that are represented in different reference states), we call this ``odd elasticity, once removed.''

In the classic continuum mechanics of prestressed solids, the constitutive response tensor $B_{ijkl}$ links the variation of Cauchy stress to the Lagrangian strain (described in more detail below).
This tensor serves as the effective stiffness that determines the material's stability and its acoustic response.
Its broad relevance can be seen in applications ranging from wave propagation in planetary cores under high pressure \cite{karki2001high} to compression stiffening in biopolymer networks \cite{engstrom2019compression}. A strikingly similar tensor also appears in a quantum Hall setting~\cite{osborne2026geometry}.

Unlike a standard elastic tensor, $B_{ijkl}$ does not have to possess major symmetry \cite{barron1965second,wallace1967thermoelasticity,biot1964mechanics,thurston1964third}: its anti-symmetric part is determined by the prestress $S_{ij}$ via 
\begin{equation}\label{eq:BAntisymmetry}
B_{ijkl} -B_{klij}= S_{kl}\delta_{ij} - S_{ij}\delta_{kl}.
\end{equation}
Although the full expressions for $B_{ijkl}$ are crucial in acoustoelastic analyses~\cite{man1987towards,duquennoy2008effective}, this anti-symmetric component is often treated as a mathematical complication.
Many studies, e.g., those reviewed in Sections II.B.3,4 of Ref.~\cite{grimvall2012lattice}, have thus focused on the simplified case of isotropic prestress ($S_{ij} \propto \delta_{ij}$), for which this anti-symmetric contribution vanishes.
We instead use this as a design tool.
By applying a specific anisotropic prestress, we show that one can drive the system to an extreme limit where the \emph{symmetric} part of the tensor vanishes, yielding a purely anti-symmetric response satisfying $B_{ijkl} = -B_{klij}$.

Tuning to this anti-symmetric state fundamentally changes the material's acoustic properties.
We demonstrate this by inverse-designing 2D lattice models whose response is tuned to satisfy $B_{ijkl} = -B_{klij}$ along with conditions needed for dynamical stability.
These models act as unusual waveguides supporting two decoupled modes along specific directions: a string-like wave whose frequency is proportional to the wavevector ($\omega \sim |q|$) and an unconventional soft mode with an in-plane but ``flexural'' character ($\omega \sim q^2$). 
Here, we first derive our design principle, then present our lattice models and analyze their exotic wave phenomena.
We further demonstrate numerically that these effects are robust beyond the linear regime, suggesting a pathway for creating unusual materials that guide mechanical waves.

In prestressed continuum elasticity~\cite{huang1950atomic,born1954dynamical,barron1965second,wallace1967thermoelasticity,wallace1970thermoelastic,wallace1972thermodynamics,didonna2005nonaffine}, the free energy $F$ relative to the free energy and volume in a reference configuration ($F_0$ and $V_0$, respectively) having mass point positions $\bm{X}$ is given by 
\begin{equation}\label{eq:epsExpansion}
\frac{F-F_0}{V_0} = S_{ij} \epsilon_{ij} +\frac{1}{2}C_{ijkl} \epsilon_{ij} \epsilon_{kl} + \dots.
\end{equation}
This expansion involves the current mass point positions, $\bm{x}$, in terms of which the displacement gradients are $u_{ij} = \partial (x_i-X_i)/\partial X_j$ and the Lagrangian strains are $\epsilon_{ij} = \frac{1}{2}(u_{ij} + u_{ji} + u_{ki}u_{kj})$.
The symmetric tensor $S_{ij}$ is the prestress -- the Cauchy stress in the reference configuration -- and $C_{ijkl}$ is the elastic modulus tensor.
While Eq.~\eqref{eq:epsExpansion} is superficially an extension of standard linear elasticity, describing the free energy density consistently to second order in $u_{ij}$ requires using the nonlinear strain tensor.
In other words, if one expands the free energy density in powers of $u_{ij}$ instead of $\epsilon_{ij}$, the coefficients of the quadratic terms are not equal to the $C_{ijkl}$, rather they are the lower symmetry ``Huang coefficients'' $S_{ijkl} = C_{ijkl} + S_{jl}\delta_{ik}$ \cite{huang1950atomic,born1954dynamical}.

Key to our work is the constitutive relation linking the variation of the current Cauchy stress, $\sigma_{ij}$, to the Lagrangian strain.
This relationship is governed by the $B_{ijkl}$ tensor, which appears in the generalized Hooke's law for perturbations around a prestressed state \cite{barron1965second,wallace1967thermoelasticity,levitas2021nonlinear}:
\begin{equation} \label{Hooke}
    \sigma_{ij} = S_{ij} + B_{ijkl}\epsilon_{kl} + \textrm{rotation-dependent terms}.
\end{equation}
The components of this effective stiffness tensor are
\begin{equation} \label{Bijkl}
    \begin{split}
        B_{ijkl} &= C_{ijkl}+ \frac{1}{2}\Bigl( S_{il}\delta_{jk}+S_{ik}\delta_{jl} + S_{jl}\delta_{ik} \\
          &\quad\quad\quad\quad\quad\quad\quad\;\;\,\,\;\;\, + S_{jk}\delta_{il} - 2S_{ij}\delta_{kl} \Bigr).
\end{split}
\end{equation}
As in \cref{eq:BAntisymmetry}, the major symmetry of $B_{ijkl}$ is controlled by the prestress, and anistropic prestress can be used to tune the system to the entirely anti-symmetric limit of $B_{ijkl} = -B_{klij}$.
The condition on the prestress for this can be expressed as a special relationship between the Huang coefficients:
\begin{equation} \label{recipe1}
S_{ijkl}+S_{jilk}+S_{kijl} +S_{iklj} - S_{ikjl} - S_{kilj} = 0. 
\end{equation}
While concise, alternate expressions can be formulated that make specific derivations more transparent (see the Supplemental Material \cite{supMat}).

While a major anti-symmetric response tensor typically suggests non-conservative physics (akin to odd elasticity), our system is strictly conservative.
This apparent tension is resolved by recognizing that $B_{ijkl}$ does not connect an \emph{energy-conjugate} stress-strain pair~\cite{hackett2018hyperelasticity}.
True conjugate pairing, such as the second Piola-Kirchoff stress~\cite{marsden1983mathematical} $\sigma_{ij}^{II}$ and the Lagrangian strain $\epsilon_{ij}$, rely on the standard elastic constants, $C_{ijkl}$.
Because our tuning leaves the major symmetry of $C_{ijkl}$ untouched, energy conservation is guaranteed.
What then, are the physical consequences of a major anti-symmetry in $B$?

The answer lies in the material's acoustic response, which is governed by the dispersion equation for a prestressed solid~\cite{huang1950atomic,born1954dynamical,barron1965second,wallace1967thermoelasticity}: $|S_{ijkl}\xi_j\xi_l - \rho_0 v^2\delta_{ik}|=0$, where $\vec{\xi}$ is a unit vector specifying the direction of wave propagation, $\rho_0$ is the density in the reference configuration, and $v$ is the wave speed.
In the purely anti-symmetric limit of $B_{ijkl} = -B_{klij}$, this equation becomes $\left|\frac{1}{2}(S_{jl}\delta_{ik} - S_{ik}\delta_{jl})\xi_j\xi_l - \rho_0 v^2\delta_{ik}\right|=0$.
Real-valued wave speeds require positive semi-definiteness of the matrix
\begin{equation}
(S_{jl}\delta_{ik} - S_{ik}\delta_{jl})\xi_j\xi_l = 
\begin{pmatrix}
S_{jl}\xi_j\xi_l - S_{xx} & -S_{xy} \\
-S_{yx} & S_{jl}\xi_j\xi_l - S_{yy}
\end{pmatrix}.
\end{equation}
For an arbitrary propagation direction in 2D, this condition simultaneously demands $S_{xy}=S_{yx}=0$, $S_{xx}-S_{yy}\geq 0$, and $S_{yy}-S_{xx}\geq 0$ -- i.e., full dynamical stability in a 2D bulk system requires either isotropic or zero prestress, both of which are incompatible with the anti-symmetric state.

However, dynamical stability can be achieved for waves propagating along special directions, effectively turning the material into a highly selective waveguide.
For instance, for propagation restricted to the $\hat{x}$ direction, the anti-symmetric response is stable  under the general tuning condition
\begin{align}
S_{xy} = S_{yx} &= 0, \label{stableRecipe0} \\
S_{xx} - S_{yy} &\geq 0, \label{stableRecipe1} \\
S_{xxyy} = S_{yyxx} = -S_{xyyx} = -S_{yxxy} &= \frac{S_{xx}+S_{yy}}{2}, \label{stableRecipe2} \\
S_{xyxy} = -S_{yxyx} &= \frac{S_{yy}-S_{xx}}{2}, \label{stableRecipe3}
\end{align}
and all other $S_{ijkl}$ set to zero.
We show in the Supplemental Material that a continuum elastic system tuned to this state supports two distinct modes \cite{supMat}.
The first is a transverse string-like wave with a linear dispersion and a speed set by the prestress anisotropy, $v_T \propto \sqrt{S_{xx} - S_{yy}}$.
The second is a longitudinal zero mode, raising the possibility of a nonlinear dispersion in a discrete system.


\begin{table}[ht]
\centering
\caption{Lattice metamaterials derived from Eq.~\eqref{simpleRecipe}. The $k$'s are the bond stiffnesses while the $k^{\textrm{eff}}$'s are the effective dynamical stiffnesses~\cite{didonna2005nonaffine} equal to the $k$'s minus the corresponding ratios of bond pretension to bond length. \label{latticeModels}}
\begin{tabular}{|c|c|c|c|c|c|c|}
\hline
Wigner-Seitz cell & $k_A$ & $k_B$ & $k_C$ & $k_A^{\textrm{eff}}$ & $k_B^{\textrm{eff}}$ & $k_C^{\textrm{eff}}$ \\
\hline
\begin{minipage}{.1\textwidth}
    \vspace{0.05cm}
    \includegraphics[width=\linewidth]{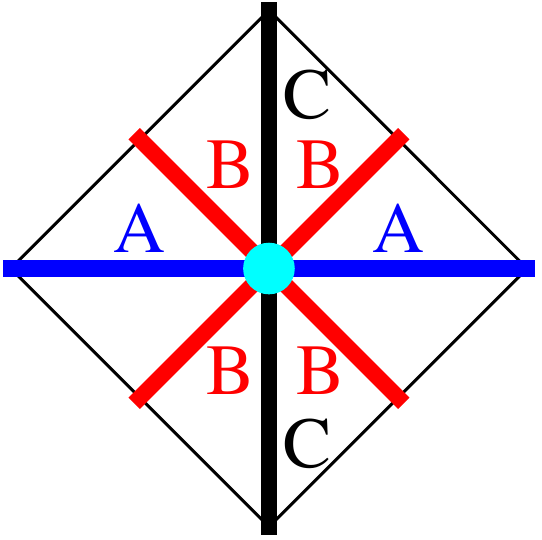}
\end{minipage}
& $\dfrac{-k_B}{2}$ & $>0$ & $\dfrac{-k_B}{2}$ & $\dfrac{-S_{xx}}{2}$ & $0$ & $\dfrac{S_{xx}}{2}$ \\
\hline
\begin{minipage}{.1\textwidth}
    \vspace{0.05cm}
    \hspace*{0.09cm}\includegraphics[width=\linewidth]{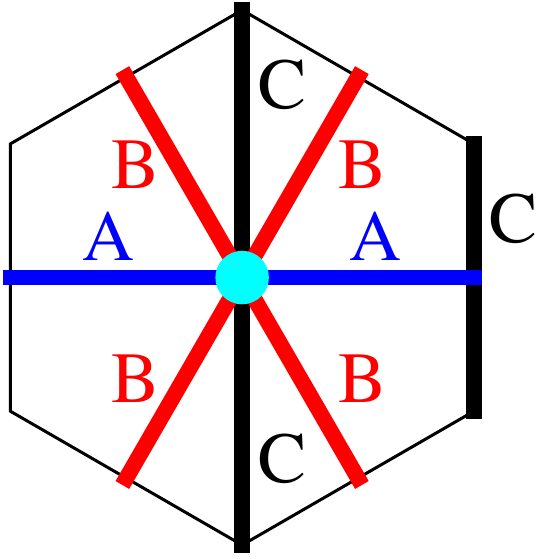}
\end{minipage}
& $\dfrac{-k_B}{2}$ & $>0$ & $\dfrac{-k_B}{2}$ & $\dfrac{-\sqrt{3}S_{xx}}{2}$ & $0$ & $\dfrac{\sqrt{3}S_{xx}}{6}$ \\
\hline
\end{tabular}
\end{table}

\begin{figure}[ht] 
\centering  
\includegraphics[width=\linewidth]{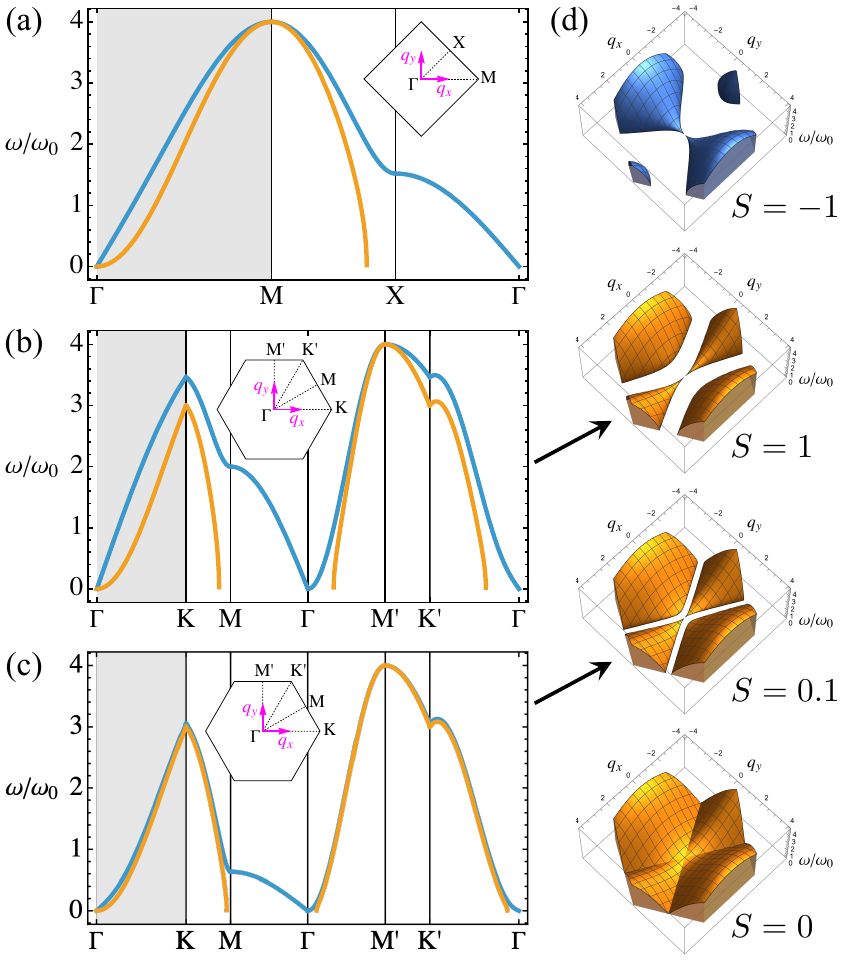}
\caption{Metamaterial phonon spectra in units of $\omega_0=\sqrt{k_B/2m}$ with $\sqrt{D_{xx}}$ shown in orange and $\sqrt{D_{yy}}$ shown in blue of (a) the $S=1$ square lattice designed for stable propagation in the shaded gray region; (b) the $S=1$ triangular lattice designed for stable propagation in the shaded gray region; (c) the $S=0.1$ triangular lattice; (d) the lower phonon branch of the triangular lattice, demonstrating how $S$ controls the stability landscape (empty regions indicate instabilities).}
\label{phonons}
\end{figure}

To demonstrate the consequences of stabilized waves propagating in special directions in these tuned materials, we inverse-design minimal lattice metamaterials that realize the waveguide physics predicted by the continuum theory.
We target simple but non-trivial cases of the general recipe in \cref{stableRecipe0,stableRecipe1,stableRecipe2,stableRecipe3}, namely
\begin{equation}
S_{yxyx}=-S_{xyxy} = S_{xx} = -S_{yy}, \label{simpleRecipe}
\end{equation}
with all other $S_{ij}$ and $S_{ijkl}$ set to zero and with $S_{xx}>0$ for stable propagation in the $\hat{x}$ direction or $S_{xx}<0$ for stable propagation in the $\hat{y}$ direction.
We solve this inverse design problem in 2D square and triangular lattices with harmonic central-force springs.
We find that the minimal solution requires both nearest- and next-nearest-neighbor-interactions using three distinct types of springs, as shown in \cref{latticeModels}.
In particular, the solutions require a combination of positive and negative stiffness bonds; we parameterize our solutions in terms of the stiffness of the ``B''-type bonds, $k_B$, and a dimensionless prestress $S\equiv \sqrt{3}S_{xx}/k_B$ (see the Supplemental Material for details \cite{supMat}).

We compute the exact phonon dispersion for these lattices, as shown in \cref{phonons} and with details given in the Supplemental Material \cite{supMat}, revealing the dramatic consequences of tuning the system to the major anti-symmetric state in this way.
Along the axis of stability, the material acts as a mechanical polarizer, supporting two decoupled modes.
The first is a transverse-polarized string-like wave with a linear dispersion, $\omega_T\sim \sqrt{S}|q|$, as expected.
The second is a longitudinal mode that transforms into an in-plane ``flexural''-like soft mode with a quadratic dispersion, $\omega_L\sim q^2$ -- a dispersion usually associated with the signature of a buckling instability.
As $S\to0$, these two modes collapse onto a common dispersion curve that is quadratic for both $\hat x$ and $\hat y$ propagation, while an instability window in the 2D Brillouin zone that is oriented along $\Gamma$-X (square lattice) or $\Gamma$-M (triangular lattice) closes and becomes a highly degenerate ground state.

The dynamics of the unusual soft mode are captured by an effective  Euler-Bernoulli-like Lagrangian density,
\begin{equation}
\mathcal{L}=\frac{1}{2}\rho\dot Q^2 - \frac{1}{2}\kappa(Q'')^2.
\end{equation}
Here $\bm Q=\bm x-\bm X$ is the displacement of nodes from their equilibrium positions, and the ratio $\kappa/\rho$ is $k_Ba^4/8m$, $k_Ba^4/32m$, and $9k_Ba^4/32m$ for the square, $S\geq0$ triangular, and $S\leq0$ triangular lattices, respectively, where $a$ is the lattice constant and $m$ is the node mass.

This model reveals the soft mode's most striking feature: its exotic energy transport.
For a Lagrangian density involving up-to-second-order gradients of the field variables $u_i$ (a ``grade 2 material'' in elastic contexts~\cite{eshelby1975elastic}), the stress-energy tensor is given by~\cite{eshelby1975elastic, engstrom2023dynamics}
\begin{equation}
T_{\mu}^{\alpha} = \bigg[\frac{\partial\mathcal L}{\partial u_{i,\alpha}} - \partial_{\beta}\bigg(\frac{\partial\mathcal L}{\partial u_{i,\alpha\beta}}\bigg)\bigg]u_{i,\mu} + \frac{\partial\mathcal L}{\partial u_{i,\alpha\beta}}u_{i,\beta\mu} - \delta^{\alpha}_{\mu}\mathcal L.
\end{equation}
Evaluated with a trial wave solution $Q=A\sin(qx-\omega t)$, the momentum density and energy current density are 
\begin{align}
    -T_x^t &= -\frac{\partial\mathcal L}{\partial\dot Q}Q' = \rho\omega q A^2\cos^2(qx-\omega t), \label{momentumDensity}\\
    T_t^x &= -\left(\frac{\partial\mathcal L}{\partial Q''}\right)'\dot{Q} + \frac{\partial\mathcal L}{\partial Q''}\left(\dot{Q}\right)' = \kappa\omega q^3 A^2. \label{energyCurrentDensity}
\end{align}
Thus, while the momentum density oscillates as expected, the energy current density is a constant stream.
This constant flux arises from the precise out-of-phase energy transport between different bond families, analogous to the interplay between shearing force and bending moment in flexural waves propagating in elastic beams.
This wave phenomena is a striking departure from conventional, covariant wave phenomena, for which $|T_t^x/T_x^t|$ is constant and equal to the squared wave speed.

To test the robustness of these phenomena and evaluate whether these effects exist for finite-amplitude waves, we perform molecular dynamics simulations of the triangular lattice model in the NVE ensemble \cite{allen2017computer}.
We measure energy and momentum fluxes through fixed planes in our simulation using the direct Irving-Kirkwood approach \cite{irving1950statistical}, and make other standard measurements of lattice wave properties.
A description of the code used to perform the simulations is given in the Supplemental Material \cite{supMat}, and the code itself can be found at Ref.~\cite{CodeRepo}.

\begin{figure}[ht] 
    \centering  
    \includegraphics[width=0.85\linewidth]{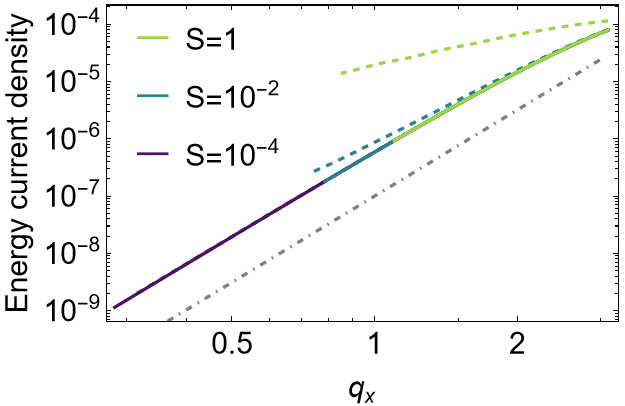}
    \caption{\label{fig:simulationTest}
        Numerical tests of predicted energy transport and stability in $S>0$ triangular lattices, from molecular dynamics simulations with finite-amplitude ($A=a/100$) waves on a $200a \times 20a$ system.
        Solid and dashed curves are longitudinal and transverse modes, respectively, with colors distinguishing the value of $S$. The dot-dashed gray line is a guide to the eye with the predicted log-log slope of 5 from Eq.~\eqref{energyCurrentDensity}.
        As $S\to0$, the increasing regime of stability of both modes can be seen by their persistence to smaller $q_x$ values, and the increasing similarity of the two modes' transport properties reflects their collapse onto a common dispersion.
    }
\end{figure}

As shown in \cref{fig:simulationTest}, the simulations confirm the existence and properties of the predicted modes in their designed-for-stability regimes. 
For propagation along $\hat{x}$, we find the long-wavelength stability of the finite-amplitude modes tracks the width of the instability window shown in \cref{phonons}(d): as $|S|$ is decreased, the window narrows, and stability improves.
Interestingly, for propagation along $\hat{y}$ (not shown in \cref{fig:simulationTest}), we find a robustly stable flexural-like mode whose measured group velocity and energy current density are also in excellent agreement with the theoretical predictions above.

The primary difference between the idealized theoretical calculations and our numerical simulations lies in the apparent stability of the long-wavelength modes.
While the theoretical dispersion along the tuned propagation axis satisfies $\omega^2 \geq 0$, the 2D lattice dynamics has off-axis sectors of instability ($\omega^2  < 0$) in the Brillouin zone near the axis (see \cref{phonons}(d)).
In our simulations, small nonlinear mode-coupling and numerical noise inevitably scatter some energy from the primary 1D mode into these unstable off-axis modes.
We have verified that these low-$q$ numerical instabilities are insensitive to boundary conditions (periodic, sliding, or longitudinally constrained), finite-size effects (over a factor of five in lattice size), and perturbation amplitudes down to near-linear perturbations ($A = 10^{-4}a$) (see Supplemental Material~\cite{supMat}).
This demonstrates that it is an intrinsic feature of the 2D energy landscape.
Reducing $|S|$ geometrically contracts and closes these off-axis instability pockets, systematically extending the dynamical stability to smaller wavevectors, as seen in~\cref{fig:simulationTest}.

Taking the limit of small $|S|$, at $S=0$ we no longer have a major anti-symmetry in $B$.
However, at this point the lattice supports circularly polarized, chiral phonons~\cite{juraschek2025chiral} in 2D, and for arbitrary wavevector.
This highly unusual property is due to the collapse of the longitudinal and transverse modes onto a common dispersion that for the triangular lattice is shown at the bottom of \cref{phonons}(d).
A simulation of such a ``spin'' wave is included in the Supplemental Material \cite{supMat}.


This work has introduced ``odd elasticity, once removed'' as a principle for metamaterial design, demonstrating how to generate a major anti-symmetric constitutive response in a conservative system.
Recent work noted that within a framework involving non-energy-conjugate stress-strain pairs, any prestressed material will appear to have odd components \cite{fruchart2023odd}.
This apparent non-conservation is connected with the distinction between general Cauchy elasticity -- which allows for responses lacking major symmetry -- and the energy-conserving framework of Green elasticity \cite{yavari2025nonlinear}.
Our key contribution is to show that tuning to generate ``odd'' components in the constitutive response of a conservative system can drive the system to a transition point: the anisotropic prestress precisely cancels the linear term in the acoustic dispersion, allowing a higher-order soft mode to dominate the long-wavelength dynamics.
In two different lattice geometries, this results in an exotic, in-plane flexural-like mode with a quadratic dispersion ($\omega\sim q^2$).

One of the more surprising results of our analysis is the apparently limited role that the magnitude of the prestress, $S$, plays in the long-wavelength physics of the soft mode.
The \emph{sign} of $S$ is crucial for stabilizing propagation in different lattice directions, and its magnitude sets the speed of the transverse wave and regime of stability.
But the longitudinal soft mode's quadratic dispersion is seemingly insensitive to it.
This suggests that once the system is tuned to the critical point, the form of the emerging soft mode is generic, rather than one that depends on the intensity of the tuning.

While we have solved a few instances of the inverse design problem posed by the recipe described above, much remains to be explored.
Extending these design principles to 3D could lead to materials with novel bulk and surface waves.
From an experimental perspective, the reliance on both positive and negative stiffness springs in our minimal models points directly toward realizations using assemblies of pre-buckled beams, post-buckling bistable trusses, or magnetic elements.
To test whether these phenomena survive in realistic nonlinear settings, we implemented simulations where negative-stiffness bonds are replaced by bistable quartic double-well potentials matched to the target curvature at equilibrium.
These simulations revealed a striking dichotomy: while the transverse mode remains stable and matches the harmonic system (including when it softens to the exotic dispersion at low $|S|$), the longitudinal soft mode has unstable low-wavevector behavior across the range of $S$ we studied.
This indicates that while the longitudinal soft mode and its associated DC energy flux rely delicately on the linearity of the negative-stiffness elements, the transverse branch at small $|S|$ will remain more stable. 
This dichotomy arises from the kinematics: longitudinal waves modulate the negative-stiffness bond lengths at first order in strain ($\Delta l \sim \partial_x Q$, readily exploring the nonlinearities in the bistable well), whereas transverse displacements couple to bond stretching only at second order ($\Delta l \sim (\partial_x Q)^2$).
Engineering metamaterial elements that can maintain nearly perfectly linear negative stiffness over finite amplitudes—perhaps through carefully designed buckling beams or reversible delamination processes—thus stands as a motivating challenge for future experimental design, and we do note some recent progress on this~\cite{song2023linear,yuan2025linear}.
Whether a combination of positive and negative incremental stiffnesses is a strictly generic requirement for conservative odd elasticity, or an artifact of the central-force lattices considered here, remains an intriguing open question.

Finally, the unique wave physics of the anti-symmetric and $S=0$ states discussed here potentially open several avenues for potential applications.
The soft mode's constant energy current is a remarkable feature.
This suggests a mechanism for creating a ``quiet'' energy channel that transports energy at a steady rate without the associated oscillations in momentum and energy density -- such a ``DC waveguide'' might enable devices that deliver vibrational energy with minimal local disturbance.
Out-of-plane angular momentum carried by circularly polarized, chiral phonons in the $S=0$ metamaterials could form the basis of novel torque actuators, or thin structures that are gyroscopically stabilized against bending.   
The unconventional dispersion involved in both of these phenomena also hints at unusual wavepacket dynamics, such as unusual spreading behavior, which warrants further investigation.

\begin{acknowledgments}
  This research was supported by Cottrell Scholar Award \#CS-CSA-2025-038 from Research Corporation for Science Advancement (TAE), and by grant NSF PHY-2309135 to the Kavli Institute for Theoretical Physics (KITP) (DMS).
  We thank Anton Souslov, Ian Tan, Chris Santangelo, Valery Levitas, and Ian Osborne for stimulating discussions.
\end{acknowledgments}
\twocolumngrid 
\bibliography{elasticityBib}
\newpage
\onecolumngrid 
\appendix

\section{Appendix A: Derivation of the prestress tuning recipe}

The starting point for tuning to the major anti-symmetric state is to set the right-hand side of Eq.~(1) equal to $-2B_{klij}$. Doing this and relabeling indices gives
\begin{equation}
S_{ij}\delta_{kl} - S_{kl}\delta_{ij} = -2B_{ijkl}.
\end{equation}
From here, substituting Eq.~(4) and using the relationship between the elastic constants and Huang coefficients ($C_{ijkl} = S_{ijkl} - S_{jl}\delta_{ik}$) yields a useful version of the prestress tuning recipe
\begin{equation}
S_{ijkl} = \frac{1}{2}\left( S_{ij}\delta_{kl} + S_{kl}\delta_{ij} + S_{jl}\delta_{ik} -S_{ik}\delta_{jl} -S_{il}\delta_{jk} - S_{jk}\delta_{il}  \right). \label{altRecipe}
\end{equation}
Further manipulations using the following two symmetry properties of the Huang coefficients (cf. the first and third lines of Eq.~(4.4) in Ref.~\cite{barron1965second})
\begin{align}
    S_{abcd} - S_{bacd}&=S_{bd}\delta_{ac} - S_{ad}\delta_{bc}, \\
    S_{abcd} - S_{badc}&=S_{bd}\delta_{ac} - S_{ac}\delta_{bd}, 
\end{align}
lead to the version given by Eq.~(5).

Making the recipe compatible with dynamical stability in a particular direction involves two auxiliary conditions that guarantee positive semidefiniteness of Eq.~(6): $S_{xy}=S_{yx}=0$ and
\begin{equation}
S_{xx}-S_{yy} \geq0, \;\;\;\;\textrm{for stable propagation along }\hat{x},
\end{equation}
or
\begin{equation}
S_{xx}-S_{yy} \leq0, \;\;\;\;\textrm{for stable propagation along }\hat{y}.
\end{equation}
Equations~(7)-(10) combine these auxiliary conditions with the result of imposing the first auxiliary condition on \cref{altRecipe}.

\section{Appendix B: Elastic wave solutions}

    \subsection{Continuum elastic solution}

We start with the dispersion equation $\left|\frac{1}{2}(S_{jl}\delta_{ik} - S_{ik}\delta_{jl})\xi_j\xi_l - \rho_0 v^2\delta_{ik}\right|=0$, which can be obtained by substituting Eq.~(4.20) into Eq.~(4.22) in Ref.~\cite{barron1965second} and then setting $B_{ijkl}=-B_{klij}$. Specializing in this section to the conditions needed for stable propagation along $\hat{x}$, we obtain
\begin{equation}
\left|
\begin{pmatrix}
0 & 0\\
0 & \frac{S_{xx}-S_{yy}}{2\rho_0}
\end{pmatrix}
- v^2\delta_{ik}\right|=0.
\end{equation}
Following Ref.~\cite{landau1986theory}, we substitute each of the two roots $v^2=0,(S_{xx}-S_{yy})/2\rho_0$ back into the Fourier-transformed equation of motion
\begin{equation}
\left[
\begin{pmatrix}
0 & 0\\
0 & \frac{S_{xx}-S_{yy}}{2\rho_0}
\end{pmatrix}
- v^2\delta_{ik}\right]\phi_k=0,
\end{equation}
where $\bm{\phi}$ is the wave amplitude, and solve for $\bm{\phi}$. This reveals that the zero mode is longitudinally polarized ($\phi=\phi_x$) and the mode with $v=\sqrt{(S_{xx}-S_{yy})/2\rho_0}$ is transversely polarized ($\phi=\phi_y$).

    \subsection{General aspects of lattice solutions}

For a Bravais lattice in which all bond energies $f$ are of a central force variety $f(\rho=r^2)$, the prestresses and Huang coefficients have analytic expressions in terms of the equilibrium bond lengths and force constants~\cite{barron1965second}:
\begin{align}
S_{ij} &= \frac{1}{2V_{\textrm{cell}}}\sum_{\bm\delta} 2\delta_i\delta_j\frac{df}{d\rho}\bigg|_{\delta^2}, \label{atomicExpression1}\\
S_{ijkl} &= S_{jl}\delta_{ik} + \frac{1}{2V_{\textrm{cell}}}\sum_{\bm\delta} 4\delta_i\delta_j\delta_k\delta_l\frac{d^2f}{d\rho^2}\bigg|_{\delta^2}. \label{atomicExpression2}
\end{align}
Here and below, the notation (after \cite{mahan2011condensed}) is such that $\bm{\delta}=\delta\hat\delta$ locate bonded neighbors of node $p$ in the prestressed reference configuration, e.g., if node $q$ is bonded to node $p$, then $\bm{\delta}=\bm X_q-\bm X_p$ for that bond. 

We consider lattices involving up to three different types of central force bonds (labeled A,B,C), each of the Hookean form
\begin{equation}
f_{A,B,C}(\rho) = \frac{1}{2}k_{A,B,C}\left(\sqrt\rho-r_{A,B,C}\right)^2. \label{springEnergy}
\end{equation}

For a given lattice geometry and bond configuration, the recipe for the dynamically stable anti-symmetric state (Eqs.~(7) to (10)) can be imposed on \cref{atomicExpression1,atomicExpression2,springEnergy}. If a solution exists and can be inverted to obtain the $k_{A,B,C}$ and $r_{A,B,C}$, the phonons may then be calculated from the equation of motion~\cite{zhang2022prestressed,didonna2005nonaffine}
\begin{equation}
m\frac{\partial^2}{\partial t^2}\bm Q_p = -\sum_{\bm\delta} \bigg{\{} \bigg[k(\bm\delta)-\frac{t(\bm\delta)}{\delta}\bigg] \hat\delta\hat\delta \cdot (\bm Q_p - \bm Q_{p+\bm\delta}) + \frac{t(\bm\delta)}{\delta}(\bm Q_p - \bm Q_{p+\bm\delta}) \bigg{\}}. \label{EOM1}
\end{equation}
Here $\bm Q_p=\bm x_p-\bm X_p$ is the displacement of node $p$ from its equilibrium position, while $k(\bm{\delta})$ and $t(\bm{\delta})=(df/dr)|_{\delta}$ are the spring constant and pretension of bond $\bm{\delta}$, respectively. The part in square brackets can be regarded as an effective spring constant for the bond-directed component of the force. Assuming the usual form of a phonon wave $\bm Q_p=\bm Q\exp[i(\bm{q\cdot X}_p-\omega t)]$, \cref{EOM1} becomes an eigenvalue equation
\begin{equation}
D \begin{pmatrix} Q_x \\ Q_y \end{pmatrix} = \omega^2 \begin{pmatrix} Q_x \\ Q_y \end{pmatrix}, \label{EOM2}
\end{equation}
where the square roots of the eigenvalues of the dynamical matrix $D$ give the phonon frequencies $\omega(\bm q)$.

    \subsection{Specific lattice solution: square lattice}

We first consider a square lattice with translation vectors $\mathbf{a}_1=(a/\sqrt{2}, a/\sqrt{2})$ and $\mathbf{a}_2=(-a/\sqrt{2}, a/\sqrt{2})$ such that $V_{\textrm{cell}}=a^2$. ``A'' bonds are defined to be second-nearest-neighbor bonds in the $x$ direction (parallel to $\mathbf{a}_1-\mathbf{a}_2$), ``B'' bonds are defined to be first-nearest-neighbor bonds, and ``C'' bonds are defined to be second-nearest-neighbor bonds in the $y$ direction (parallel to $\mathbf{a}_1+\mathbf{a}_2$). \cref{atomicExpression1,atomicExpression2} become, after simplification,
\begin{align}
S_{xx}&= 2k_A(1-r_A/\sqrt{2}a) + k_B(1-r_B/a),\\
S_{yy}&= 2k_C(1-r_C/\sqrt{2}a) + k_B(1-r_B/a),\\
S_{xy}=S_{yx}&= 0,\\
S_{xxxx}&= S_{xx} + \sqrt{2}k_Ar_A/a + k_Br_B/2a,\\
S_{yyyy}&= S_{yy} + \sqrt{2}k_Cr_C/a + k_Br_B/2a,\\
S_{xxxy}=S_{xxyx}=S_{xyxx}=S_{yxxx}&=0,\\
S_{yyyx}=S_{yyxy}=S_{yxyy}=S_{xyyy}&=0,\\
S_{xxyy}=S_{yyxx}=S_{xyyx}=S_{yxxy}&=k_Br_B/2a,\\
S_{yxyx}&= S_{xx} + k_Br_B/2a, \\
S_{xyxy}&=S_{yy} + k_Br_B/2a.
\end{align}
Combining with the version of Eq.~(11) that gives stable propagation in the $x$ direction, we find
\begin{align}
k_A&=k_C=-k_B/2, \label{sqModelks}\\
r_A&=-r_C=\sqrt{2}S_{xx}a/k_B,\;\;\;\;\;r_B=0, \label{sqModelrs}\\
S_{xx}&>0. \label{sqModelSxx}
\end{align}

To facilitate the phonon calculation, we introduce notation $s=S_{xx}/k_B$, $\theta_x=\sqrt2q_xa$, and $\theta_y=\sqrt2q_ya$. Putting the square lattice model into \cref{EOM1}, dropping the subscript $p$ for brevity, and simplifying, we obtain the equation of motion
\begin{align}
-m\omega^2\mathbf{Q}  
&=2k_B\left(\mathbf{Q}-sQ_y\hat y\right)\sin^2\left(\frac{\theta_x}{2}\right)\nonumber \\
&+2k_B\left(\mathbf{Q}+sQ_x\hat x\right)\sin^2\left(\frac{\theta_y}{2}\right) - 4k_B\mathbf{Q}\left[1-\cos\left(\frac{\theta_x}{2}\right)\cos\left(\frac{\theta_y}{2}\right)\right],
\end{align}
which can be expressed in terms of dynamical matrix elements
\begin{align}
D_{xx} &= \frac{2k_B}{m}\left[-s_x^2 + 2(1-c_xc_y) - (1+s)s_y^2\right],\\
D_{yy} &= \frac{2k_B}{m}\left[-s_y^2 + 2(1-c_xc_y) - (1-s)s_x^2\right],\\
D_{xy} &= D_{yx} = 0.
\end{align}
Here the notation is $s_x=\sin(\theta_x/2)$, $c_x=\cos(\theta_x/2)$, $s_y=\sin(\theta_y/2)$, and $c_y=\cos(\theta_y/2)$. For propagation along $\hat x$, $s_y=0$ and $c_y=1$, giving
\begin{align}
\omega_L &= \sqrt{\frac{2k_B}{m}\left[-s_x^2 + 2(1-c_x) \right]} \approx \frac{1}{2}\sqrt{\frac{k_B}{2m}}(qa)^2  \;\;\;\;\;\textrm{as } q\to0,\\
\omega_T &= \sqrt{\frac{2k_B}{m}\left[(s-1)s_x^2 + 2(1-c_x) \right]} \approx \sqrt{2s}\sqrt{\frac{k_B}{2m}}|qa| = \sqrt{\frac{S_{xx}-S_{yy}}{2\rho_0}}|q|  \;\;\;\;\;\textrm{as } q\to0,
\end{align}
in agreement with the continuum elastic solution. To make contact between this lattice model and the effective Lagrangian for the longitudinal mode (Eq.~(12)), we equate coefficients in the respective quadratic dispersions to obtain the Lagrangian parameter
\begin{equation}
\frac{\kappa}{\rho} = \frac{k_Ba^4}{8m}.
\end{equation}

    \subsection{Specific lattice solution: triangular lattice}

We follow a similar path in analyzing a triangular lattice with translation vectors $\mathbf{a}_1=(a,0)$ and $\mathbf{a}_2=(a/2,\sqrt{3}a/2)$, such that $V_{\textrm{cell}}=\sqrt{3}a^2/2$. ``A'' bonds are defined to be nearest-neighbor bonds in the $x$ direction (parallel to $\mathbf{a}_1$), ``B'' bonds are defined to be nearest-neighbor bonds parallel to both $\mathbf{a}_2$ and $\mathbf{a}_2-\mathbf{a}_1$, and ``C'' bonds are defined to be second-nearest-neighbor bonds in the $y$ direction (parallel to $2\mathbf{a}_2-\mathbf{a}_1$). \cref{atomicExpression1,atomicExpression2} become, after simplification,
\begin{align}
S_{xx}&= (2k_A/\sqrt{3})(1-r_A/a) + (k_B/\sqrt{3})(1-r_B/a),\\
S_{yy}&= \sqrt{3}k_B(1-r_B/a) + 2\sqrt{3}k_C(1-r_C/\sqrt{3}a),\\
S_{xy}=S_{yx}&= 0,\\
S_{xxxx}&= S_{xx} + 2k_Ar_A/\sqrt{3}a + k_Br_B/4\sqrt{3}a,\\
S_{yyyy}&= S_{yy} + 9k_Br_B/4\sqrt{3}a + 2k_Cr_C/a,\\
S_{xxxy}=S_{xxyx}=S_{xyxx}=S_{yxxx}&=0,\\
S_{yyyx}=S_{yyxy}=S_{yxyy}=S_{xyyy}&=0,\\
S_{xxyy}=S_{yyxx}=S_{xyyx}=S_{yxxy}&= \sqrt{3}k_Br_B/4a,\\
S_{yxyx}&= S_{xx} + \sqrt{3}k_Br_B/4a, \\
S_{xyxy}&=S_{yy} + \sqrt{3}k_Br_B/4a.
\end{align}
Combining with Eq.~(11), we find
\begin{align}
k_A&=k_C=-k_B/2, \\
r_A&=-\sqrt{3}r_C=\sqrt{3}S_{xx}a/k_B,\;\;\;\;\;r_B=0, \\
S_{xx}&>0, \;\;\;\;\textrm{for stable propagation along }\hat{x},\textrm{ or} \\
S_{xx}&<0, \;\;\;\;\textrm{for stable propagation along }\hat{y}.
\end{align}

For the phonon calculation, we introduce  $S=\sqrt{3}S_{xx}/k_B$, $\theta_x=q_xa$, and $\theta_y=q_ya$. Putting the triangular lattice model into \cref{EOM1}, dropping the subscript $p$ for brevity, and simplifying, we obtain the equation of motion
\begin{align}
-m\omega^2\mathbf{Q}  
&=2k_B\left(\mathbf{Q}-SQ_y\hat y\right)\sin^2\left(\frac{\theta_x}{2}\right)\nonumber \\
&+2k_B\left(\mathbf{Q}+\frac{S}{3}Q_x\hat x\right)\sin^2\left(\frac{\sqrt3\theta_y}{2}\right) - 4k_B\mathbf{Q}\left[1-\cos\left(\frac{\theta_x}{2}\right)\cos\left(\frac{\sqrt3\theta_y}{2}\right)\right],
\end{align}
which can be expressed in terms of dynamical matrix elements
\begin{align}
D_{xx} &= \frac{2k_B}{m}\left[-s_x^2 + 2(1-c_xc_y) - (1+S/3)s_y^2\right],\\
D_{yy} &= \frac{2k_B}{m}\left[-s_y^2 + 2(1-c_xc_y) - (1-S)s_x^2\right],\\
D_{xy} &= D_{yx} = 0.
\end{align}
We define the auxiliary variables  $s_x=\sin(\theta_x/2)$, $c_x=\cos(\theta_x/2)$, $s_y=\sin(\sqrt3\theta_y/2)$, and $c_y=\cos(\sqrt3\theta_y/2)$. For propagation along $\hat x$, $s_y=0$ and $c_y=1$, giving
\begin{align}
\omega_L &= \sqrt{\frac{2k_B}{m}\left[-s_x^2 + 2(1-c_x) \right]} \approx \frac{1}{4}\sqrt{\frac{k_B}{2m}}(qa)^2  \;\;\;\;\;\textrm{as } q\to0,\\
\omega_T &= \sqrt{\frac{2k_B}{m}\left[(S-1)s_x^2 + 2(1-c_x) \right]} \approx \sqrt{S}\sqrt{\frac{k_B}{2m}}|qa| = \sqrt{\frac{S_{xx}-S_{yy}}{2\rho_0}}|q|  \;\;\;\;\;\textrm{as } q\to0,
\end{align}
and for propagation along $\hat{y}$, $s_x=0$ and $c_x=1$, giving
\begin{align}
\omega_L &= \sqrt{\frac{2k_B}{m}\left[-s_y^2 + 2(1-c_y) \right]} \approx \frac{3}{4}\sqrt{\frac{k_B}{2m}}(qa)^2  \;\;\;\;\;\textrm{as } q\to0,\\
\omega_T &= \sqrt{\frac{2k_B}{m}\left[-(1+S/3)s_y^2 + 2(1-c_y) \right]} \approx \sqrt{-S}\sqrt{\frac{k_B}{2m}}|qa| = \sqrt{\frac{S_{yy}-S_{xx}}{2\rho_0}}|q|  \;\;\;\;\;\textrm{as } q\to0,
\end{align}
all in agreement with the continuum elastic solution. To make contact between this lattice model and the effective Lagrangian for the longitudinal mode (Eq.~(12)), we equate coefficients in the respective quadratic dispersions to obtain the Lagrangian parameter
\begin{equation}
\frac{\kappa}{\rho} = 
\begin{cases}
k_Ba^4/32m, \;\;\;\;\textrm{for propagation along } \hat{x},\\
9k_Ba^4/32m, \;\;\;\;\textrm{for propagation along } \hat{y}.
\end{cases}
\end{equation}

\section{Appendix C: Numerical methods and simulation details}

\subsection{System setup and interparticle interactions}

To evaluate wave propagation beyond the infinitesimal-amplitude linear regime, we perform microcanonical ($NVE$) molecular dynamics simulations of discrete triangular lattices.
The system comprises $N = N_x \times N_y$ particles of uniform mass $m \equiv 1$ arrayed in a simulation box of dimensions $L_x = N_x a$ and $L_y = \frac{\sqrt{3}}{2} N_y a$, where $a \equiv 1$ is the lattice constant.
Unless otherwise specified, we choose $N_x = 200$ and $N_y = 20$ for wave propagation along $\hat{x}$, and $N_x = 20$ and $N_y = 200$ for propagation along $\hat{y}$; periodic boundary conditions are enforced via the usual minimum image convention.

The potential energy of the lattice is governed by central-force bonds:
\begin{equation}
V = \sum_{\langle i, j \rangle} f_{ij}(r_{ij}),
\end{equation}
where $r_{ij} = |\mathbf{r}_j - \mathbf{r}_i|$ is the instantaneous distance between bonded particles $i$ and $j$.
In our baseline harmonic simulations, each bond interacts through a Hookean potential:
\begin{equation}
f(r) = \frac{1}{2} k \left(r - r_0\right)^2,
\end{equation}
exerting a pair force on particle $i$ given by
\begin{equation}
\mathbf{F}_{ij} = -\nabla_{\mathbf{r}_i} f(r_{ij}) = k \left(r_{ij} - r_0\right) \frac{\mathbf{r}_j - \mathbf{r}_i}{r_{ij}}.
\end{equation}
We set $k_B \equiv 1$ as our characteristic unit of stiffness, giving a natural unit of time $\tau_0 = \sqrt{m / k_B} = 1$, and study bond stiffnesses ($k_A, k_B, k_C$) and rest lengths ($r_A, r_B, r_C$) as described in the main text and above.

We have also implemented as part of the repository the ability to study lattices with bistable quartic springs whose double-well potential takes the form
\begin{equation}
f_{\text{bistable}}(r) = k_4 (r - r_0)^4 - k_2 (r - r_0)^2,
\end{equation}
with $k_2 = 2U / \delta_w^2$ and $k_4 = U / \delta_w^4$, yielding two local energy minima of depth $U$ separated by $\pm \delta_w$ around a central barrier at $r_0$.
The corresponding force is
\begin{equation}
\mathbf{F}_{ij} = \left[ 4 k_4 (r_{ij} - r_0)^3 - 2 k_2 (r_{ij} - r_0) \right] \frac{\mathbf{r}_j - \mathbf{r}_i}{r_{ij}}.
\end{equation}
For a desired well depth $U$, we have solved for the parameters $(r_0, \delta_w)$ that match the effective stiffness $K$ and force $F$ of the linear target spring at the reference distance $L$.

As discussed in the main text, while the idealized harmonic negative-stiffness springs yield the exotic wave phenomena presented, standard physical realizations often involve geometric nonlinearities.
Testing the lattice with these bistable quartic double-well potentials reveals a  dichotomy between the two supported wave modes.
The transverse mode remains robust and its behavior closely matches that of the perfectly harmonic system, including exhibiting systematically improved stability at low wavevectors as $|S| \to 0$.
However, the longitudinal soft mode exhibits catastrophic instabilities for all tested values of $S$ when using this generic nonlinear implementation.
These numerical tests demonstrate that the quadratic soft-mode dispersion and the steady ``DC'' energy current delicately depend on the strict linearity of the negative-stiffness elements over finite wave amplitudes.

\subsection{Time integration and wave initialization}

We implement the Velocity Verlet algorithm using the standard three-step update scheme:
\begin{align}
\mathbf{v}_i\left(t + \frac{\Delta t}{2}\right) &= \mathbf{v}_i(t) + \frac{1}{2}\mathbf{a}_i(t)\Delta t, \\
\mathbf{r}_i(t + \Delta t) &= \mathbf{r}_i(t) + \mathbf{v}_i\left(t + \frac{\Delta t}{2}\right)\Delta t, \\
\mathbf{v}_i(t + \Delta t) &= \mathbf{v}_i\left(t + \frac{\Delta t}{2}\right) + \frac{1}{2}\mathbf{a}_i(t + \Delta t)\Delta t,
\end{align}
where $\mathbf{a+\Delta t}_i(t) = \frac{1}{m} \sum_{j} \mathbf{F}_{ij}(t + \Delta t)$ is recomputed from the updated positions $\mathbf{r}_i(t + \Delta t)$ between the second and third steps.
We employ a timestep of $\Delta t = 10^{-3}\,\tau_0$ (with convergence checks performed down to $\Delta t = 10^{-4}\,\tau_0$).

Traveling waves are initialized by displacing the particles from their reference equilibrium positions $\mathbf{X}_i$, $\mathbf{Q}_i = \mathbf{r}_i - \mathbf{X}_i$, and assigning corresponding initial velocities:
\begin{align}
\mathbf{r}_i(0) &= \mathbf{X}_i + \bm{\mathcal{A}} \cos(\mathbf{q} \cdot \mathbf{X}_i), \\
\mathbf{v}_i(0) &= \omega_{\text{th}} \bm{\mathcal{A}} \sin(\mathbf{q} \cdot \mathbf{X}_i),
\end{align}
where $\mathbf{q} = (2\pi m_x / L_x, 2\pi m_y / L_y)$ for integer mode numbers $m_x, m_y$, and $\omega_{\text{th}}(\mathbf{q})$ is the theoretical frequency calculated from the dynamical matrix.
For longitudinal modes propagating along $\hat{x}$, we set $\bm{\mathcal{A}} = (\mathcal{A}, 0)$ with $\mathcal{A} = 10^{-2}a$ (or $\mathcal{A} = 10^{-4}a$ in low-amplitude sweeps), while for transverse modes $\bm{\mathcal{A}} = (0, \mathcal{A})$.
Measurements are collected over a duration of $t_{\text{meas}} = 4 \times (2\pi / \omega_{\text{th}})$, averaging over four full wave periods.

\subsection{Microscopic flux, energy, and pseudomomentum measurements}

The instantaneous microscopic energy current crossing a line perpendicular to axis $\alpha \in \{x, y\}$ at position $x_\alpha = L_\alpha / 2 + 0.25$ is evaluated using the Irving-Kirkwood formulation:
\begin{equation}
J_{E, \alpha}(t) = \sum_{\langle i, j \rangle \in \mathcal{C}_\alpha} \frac{1}{2} \left[ \mathbf{F}_{ij} \cdot (\mathbf{v}_i + \mathbf{v}_j) \right] \operatorname{sgn}(X_{j,\alpha} - X_{i,\alpha}),
\end{equation}
where $\mathcal{C}_\alpha$ denotes the set of bonds intersecting the plane.
The line-averaged energy current density is defined as
\begin{equation}
j_{E, x} = \frac{\langle J_{E, x} \rangle_t}{L_y}, \qquad j_{E, y} = \frac{\langle J_{E, y} \rangle_t}{L_x},
\end{equation}
where $\langle \cdot \rangle_t$ denotes the temporal average over $t_{\text{meas}}$.

The wave energy density $\epsilon$ is computed by subtracting the prestressed reference ground-state energy density from the instantaneous total energy density:
\begin{equation}
\epsilon = \frac{E_{\text{total}} - E_{\text{ground}}}{L_x L_y},
\end{equation}
where $E_{\text{total}} = \sum_i \frac{1}{2} m v_i^2 + V$.
The simulated group velocity is then extracted via the transport relation
\begin{equation}
\mathbf{v}_g = \frac{\mathbf{j}_E}{\epsilon}.
\end{equation}

To track wave momentum and pseudomomentum density, we evaluate local discrete deformation gradients.
For each node $i$, with neighbor separations $\Delta \mathbf{X}_{ij} = \mathbf{X}_j - \mathbf{X}_i$ and $\Delta \mathbf{r}_{ij} = \mathbf{r}_j - \mathbf{r}_i$, we define the reference matrix $\mathbf{B}_i = \sum_{j \in \mathcal{N}(i)} \Delta \mathbf{X}_{ij} \otimes \Delta \mathbf{X}_{ij}$ and the instantaneous deformed covariance $\mathbf{C}_i = \sum_{j \in \mathcal{N}(i)} \Delta \mathbf{r}_{ij} \otimes \Delta \mathbf{X}_{ij}$.
The local deformation gradient tensor is computed as $\mathbf{F}_i = \mathbf{C}_i \mathbf{B}_i^{-1}$.
The pseudomomentum density $\mathbf{p}_{\text{pseudo}}$ is then obtained by summing over all particles:
\begin{equation}
\mathbf{p}_{\text{pseudo}} = \frac{1}{L_x L_y} \sum_{i=1}^N \mathbf{F}_i^T (m \mathbf{v}_i).
\end{equation}

\section{Appendix D: Description of movie}

\texttt{spin\_wave\_S0.0.mp4} shows a stable, $\hat x$-propagating, right-circularly polarized wave in a $S=0$, $60a \times 20a$ triangular lattice, following its initialization from $\mathbf Q_i = \mathcal{A}[\cos(\mathbf{q\cdot X}_i)\hat x + \sin(\mathbf{q\cdot X}_i)\hat y]$ and $\dot{\mathbf Q}_i = \omega_{\textrm{th}}\mathcal{A}[\sin(\mathbf{q\cdot X}_i)\hat x - \cos(\mathbf{q\cdot X}_i)\hat y]$. $\mathcal A=0.2a$ is used in the simulation. Following Ref.~\cite{juraschek2025chiral}, we characterize this phonon mode as ``false chiral'' since it is transformed into a $-\hat x$-propagating, left-circularly polarized wave by a time-reversal operation.

\end{document}